\documentclass[aps,pre,twocolumn,superscriptaddress,longbibliography,nofootinbib,showkeys,showpacs]{revtex4-2}

\usepackage[utf8]{inputenc}
\usepackage{amsfonts}
\usepackage{amsmath}
\usepackage{appendix}
\usepackage{bm}
\usepackage{braket}
\usepackage{color}
\usepackage{comment}
\usepackage{enumerate}
\usepackage[top=2cm,left=2cm,right=2cm,bottom=2cm]{geometry}
\usepackage{graphicx}
\usepackage{hyperref}
\usepackage{mathptmx} 
\usepackage{scalerel}
\usepackage{subcaption}
\usepackage{tikz}
\usetikzlibrary{shapes.geometric, arrows}
\usepackage[utf8]{inputenc}
\begin{document}
\title{Quantum annealing inspired algorithms for the NISQ Era}

\author{Rijul Sachdeva}
\affiliation{J\"ulich Supercomputing Centre,\\
Forschungszentrum J\"ulich, D-52425 J\"ulich, Germany}
\affiliation{RWTH Aachen University, 52056 Aachen, Germany}
\author{Vrinda Mehta}
\affiliation{J\"ulich Supercomputing Centre,\\
Forschungszentrum J\"ulich, D-52425 J\"ulich, Germany}\
\author{Manpreet Singh Jattana}
\affiliation{Modular Supercomputing and Quantum Computing, Goethe University
Frankfurt,\\ Kettenhofweg 139, Frankfurt am Main, 60325, Germany}
\author{Kristel Michielsen}
\affiliation{J\"ulich Supercomputing Centre,\\
Forschungszentrum J\"ulich, D-52425 J\"ulich, Germany}
\affiliation{Department of Computer Science, University of Cologne, 50931 Cologne, Germany}
\author{Fengping Jin}
\affiliation{J\"ulich Supercomputing Centre,\\
Forschungszentrum J\"ulich, D-52425 J\"ulich, Germany}

\date{\today}

\begin{abstract}
    We study algorithms inspired by quantum annealing that are suited for the NISQ era. First, we analyze approximate quantum annealing (AQA), which employs a discretized annealing ansatz in which the time step and the number of layers are allowed to deviate from a faithful implementation of quantum annealing. Parameter scans identify regimes that reproduce annealing-like behavior with reduced resources, making them more suitable for NISQ devices. The resulting parameters can then be used as an effective warm start for the quantum approximate optimization algorithm (QAOA), improving its performance compared to random initializations. We also introduce evolving Hamiltonian quantum optimization (EHQO), a multistep variational scheme that guides the optimization process through intermediate Hamiltonians derived from the standard annealing Hamiltonian. Numerical simulations on sets of hard 2-SAT instances suggest that quantum annealing-inspired algorithms provide practical strategies for enhancing variational quantum optimization.
\end{abstract}

\maketitle

\section{Introduction}

Optimization problems lie at the centre of science and engineering, spanning diverse applications such as resource allocation, logistics, scheduling, and medicine~\cite{wang2012design,tran2016hybrid,solenov2018potential,Orus2019,Woerner2019,venturelli2019reverse,Feld_2019,sakuma2020application,Mohammadbagherpoor:2021zqa,inoue2021traffic,Micheletti_21,Bova2021,cordier2022biology}. Many of these can be expressed as large-scale combinatorial optimization problems. However, as the size and complexity of these problems grow, exact methods quickly become computationally impractical, making them unsuitable for practically relevant problem scales. As a result, we have to rely on heuristic approaches to obtain high-quality solutions within a reasonable time.

A promising candidate in this direction is quantum annealing (QA)~\cite{apolloni_quantum_1989,finnila_quantum_1994,kadowaki_quantum_1998,farhi_quantum_2000,farhi_quantum_2001}, which can be viewed as the quantum analogue of simulated annealing~\cite{kirkpatrick_optimization_1983,aarts1988simulated}, a classical meta-heuristic algorithm. In simulated annealing, thermal fluctuations enable the system to explore the energy landscape, while in QA, this role is played by quantum fluctuations. It is argued that quantum effects, specifically, quantum tunneling, can help the system escape local minima more efficiently in scenarios where the energy barriers are high and narrow. QA is based on the adiabatic theorem, which states that a system initialized in its ground state will remain in the instantaneous ground state if evolved sufficiently slowly, provided that there is no gap closing between the instantaneous ground and first excited states. This ensures that the optimal solution is reached if the ground state of the final Hamiltonian, $H_P$, correctly encodes the solution to an optimization problem. Typically, the combinatorial optimization problem can be formulated as an Ising Hamiltonian, i.e.,
\begin{equation}
    H_P = -\sum_i^N h_i \sigma_z^i - \sum_{\left <i,j\right >} J_{ij} \sigma_z^i \sigma_z^j,
    \label{eq:H_P}
\end{equation}
where $\sigma_z$ is the Pauli-z operator and $h_i$ and $J_{i,j}$ are the field and coupling constants, respectively. 

The QA process begins with a simple initial Hamiltonian, $H_I$, whose ground state is known and can be easily prepared. $H_I$ is typically chosen to be
\begin{equation}
   H_I = -\sum_{i=1}^N \sigma_x^i,
    \label{eq:H_I}
\end{equation}
where $\sigma_x$ is the Pauli-x operator.
The system is initialized in the ground state of $H_I$, i.e., the uniform superposition state $\ket{+}^{\otimes N}$, and is slowly evolved towards $H_P$ according to
\begin{equation}
    H(s) = A(s) H_I + B(s) H_P,
    \label{eq:qa-h}
\end{equation}
where $s=t/t_a$, $t_a$ is the total annealing time, and $A(s)$ and $B(s)$ are functions that control the annealing schedule chosen such that $A(0) \gg B(0)$ and $A(1) \ll B(1)$. The evolution of the state describing the system is governed by the time-dependent Schrödinger equation (TDSE)
\begin{equation}
    i \hbar \frac{\partial}{\partial t} \ket{\Psi(t)} 
    = H (t)\ket{\Psi(t)},
\end{equation}
where $\hbar$ is set to $1$ in the following. The formal solution of the TDSE is given by
\begin{equation}
    \ket{\Psi(t)} = \mathcal{T} \exp \left ({-i \int_0^t H(t) dt} \right ) \ket{\Psi(0)},
    \label{eq:exactsolution}
\end{equation}
where $\mathcal{T}$ is the time-ordering operator. 

To implement QA on classical computers, one needs to find a good approximation to Eq.~(\ref{eq:exactsolution}). One of the approaches is to use the Suzuki-Trotter product formula algorithm~\cite{suzuki_generalized_1976,suzuki_decomposition_1985,de_raedt_product_1987,huyghebaert1990product, suzuki1993}. For the Hamiltonian given by Eq.~(\ref{eq:qa-h}), the first-order approximation to Eq.~(\ref{eq:exactsolution}) reads
\begin{equation}
    \ket{\Psi(t)} \approx  \prod^p_{j=1} e^{-i \tau A(j/p)H_I} e^{-i \tau B(j/p)H_P}  \ket{+}^{\otimes N},
    \label{eq:discretized_qa}
\end{equation}
where $t=\tau p$.
See~\cite{huyghebaert1990product, suzuki1993} for higher-order product formulas. 

Due to the exponentially growing dimension of the Hilbert space, the classical simulation for solving the TDSE is strongly limited by classical computational resources. For example, storing the full quantum state of a 30 (50) qubit system using double precision floating numbers requires 16~GB (16~PB) of memory. This limitation does not arise in the alternative paradigm of quantum computing, where the algorithm is implemented directly on a gate-based quantum computer. In this paradigm, Eq.~(\ref{eq:discretized_qa}) is mapped onto a quantum circuit comprising only one- and two-qubit gates~\cite{nielsen00}. Thus, QA provides a conceptually elegant framework that, given sufficiently coherent and well-controlled quantum hardware, could potentially be leveraged to tackle a wide range of optimization problems.

However, the practical landscape of quantum computing is currently defined by the noisy intermediate-scale quantum (NISQ) era. Current quantum processors contain qubits that are imperfect along with limited coherence times, high error rates, and connectivity constraints. Such noise and hardware limitations prevent the direct realization of 
long and precise circuits required for a faithful implementation of Eq.~(\ref{eq:discretized_qa}). 

This gap between theory and implementation motivates the exploration of alternative strategies. Rather than relying on a faithful implementation of QA dynamics, it is worthwhile to consider QA-inspired hybrid approaches that can be tailored to the constraints of NISQ hardware. These methods should aim to preserve some essence of QA while reducing circuit depth, increasing tolerance to noise, and making the algorithms more practical for these near-term devices. In this context, this paper discusses three methods that follow a QA-inspired evolution to approximate the ground state of $H_P$ with reasonable probability by relaxing the requirement of strictly following the full QA trajectory. A central motivation for this study is to investigate whether good performance can be achieved without incurring the full computational cost or resource requirements associated with QA.

The first algorithm is a straightforward extension of Eq.~(\ref{eq:discretized_qa}), which we refer to as approximate quantum annealing (AQA) \cite{willsch_gpu-accelerated_2022}. Here, the parameters $\tau$ and $p$ are treated as tunable quantities chosen to minimize the energy expectation for $H_P$.

Another variational approach that has the same structure as Eq.~(\ref{eq:discretized_qa}) is the well-studied quantum approximate optimization algorithm (QAOA)~\cite{farhi_quantum_2014}. In contrast to AQA, where only two parameters ($\tau$ and $p$) are optimized, in QAOA, one relaxes parameters $\tau A(j/p)$ and $\tau B(j/p)$ into two variational parameters $\beta_j$ and $\gamma_j$ in each layer , respectively. The optimal values of these parameters are obtained through classical optimizers by minimizing the energy expectation of $H_P$. The resulting evolution may or may not bear resemblance to the quantum annealing trajectory. As for any variational algorithm, the performance of QAOA is highly sensitive to the number of layers (parameters), the initial parameter values, and the optimization method. In particular, complex energy landscapes, poorly chosen initial parameters, and vanishing gradients can significantly hinder the optimizer and, consequently, the overall performance of the algorithm.

Finally, we propose an algorithm called evolving Hamiltonian quantum optimization (EHQO), which aims to gradually traverse from an initial Hamiltonian to the problem Hamiltonian through a sequence of intermediate Hamiltonians that follow a quantum-annealing-like evolution of the instantaneous Hamiltonian [see Eq.~(\ref{eq:qa-h})]. More specifically, for a linear scheme of interpolation between the initial and the problem Hamiltonians, we define  $N_s-1$ intermediate Hamiltonians as
\begin{align}
H(s_j) = (1-s_j) H_I + s_j H_P,\nonumber\\
s_j = \frac{j}{N_s},\;\; j = 1,\cdots,N_s-1.
\label{eq:EHQO_intermediate}
\end{align}
The algorithm proceeds through $N_s$ steps. At each step $j$, a parameterized quantum circuit (to be specified later) is optimized with respect to the cost Hamiltonian $H(s_j)$. Once the optimization is completed, the resulting parameter set is passed forward and used as the initialization for the next step, which optimizes with respect to $H(s_{j+1})$. Since successive Hamiltonians differ only slightly, the optimized parameters from step $j$ typically produce a state with energy close to the ground state energy of $H(s_{j+1})$. This smooth progression helps guide the optimization process and can ease the difficulty of searching directly according to the expectation value of problem Hamiltonian $H_P$. However, it is possible that around the critical region of an avoided level crossing with respect to the QA Hamiltonian, the corresponding energy spectrum changes more significantly, making the optimization more challenging. Nevertheless, the stepwise evolution provides a structured pathway that can substantially improve the performance of variational optimization. Such approaches and other combinations of adiabatic optimization and variational algorithms have been explored before as well \cite{mcclean_theory_2016,wecker_progress_2015,farhi_quantum_2014,garcia-saez_addressing_2018,willsch_benchmarking_2020,manpreet23}.

The remainder of the paper is organized as follows. In Sections~\ref{sec:AQA}, \ref{sec:QAOA}, and \ref{sec:EHQO}, we present the basic ideas behind the AQA, QAOA, and EHQO algorithms, respectively, and discuss insights obtained from a mechanistic analysis of their dynamics. In Section~\ref{sec:PerformanceComparison}, we provide a comparative analysis of the performance of these algorithms. Finally, Section~\ref{sec:conclusion} summarizes the main conclusions of this work.

\section{Approximate Quantum Annealing}
\label{sec:AQA}
\subsection{Basic idea}
In a faithful implementation of quantum annealing (QA) [Eq.~(\ref{eq:discretized_qa})], i.e., when a sufficiently small time step $\tau$ is chosen, one can obtain high-quality solutions for large annealing times $t_a=\tau p$, implemented by choosing a large $p$. However, in the NISQ era, implementing such long evolutions on gate-based quantum devices is challenging due to the large circuit depths required.

To address this limitation, the idea behind AQA is to treat both $\tau$ and $p$ as free parameters. This relaxes the requirement that the evolution must strictly follow the QA trajectory and instead allows for deviations from the QA path. The goal is to identify parameter choices that still yield good solutions at reduced circuit depth.

From this perspective, for a general problem, one can anticipate different regimes of behaviour depending on the choice of $\tau$. For sufficiently small $\tau$, the dynamics are expected to closely follow the QA evolution. As $\tau$ increases, the trajectory departs from the QA path while still retaining some qualitative features of QA. For larger $\tau$, however, the circuit dynamics are no longer expected to maintain any correspondence with QA. In the following section, these behaviors will be explored for a specific problem.

\subsection{Results}
\label{sec:aqa_results}
We employ an instance of an 8-variable 2-SAT problem with a unique known ground state~\cite{thomas_quantum_2014}. Since there are only two free parameters and the size of the problem allows for it, here we illustrate the parameter-dependence of AQA through a scan of parameters $\tau \in [0.05,1.5]$ and $p \in [10,1000]$ for a linear schedule with $A(j/p)=1-j/p$ and $B(j/p)=j/p$.

Figure~\ref{fig:aqa-1} presents the resulting two-dimensional scan of the success probability as a function of $\tau$ and $p$ for the 8-variable 2-SAT problem, from which the three operational regimes can be identified. The first regime corresponds to approximately $\tau < 0.4$, where the contours of the success probability closely follow the red-dashed lines representing constant values of the product $\tau p$, i.e., a constant annealing time. This indicates that, in this regime, the quantum circuit faithfully reproduces the QA dynamics, as the value of the success probability stays similar to that obtained for small values of $\tau$. As shown in the figure, the success probability in this region increases with increasing $p$ for a fixed $\tau$, and with increasing $\tau$ for a fixed $p$, i.e., with increasing annealing time.

Upon further increasing $\tau$, the success probability contours start to deviate from the red lines representing constant $\tau p$, although for a constant $p$, the success probability increases until $\tau \approx 0.9$. This is the regime where the error involved in the Trotterization of the time evolution under QA becomes too large for the quantum circuit to implement QA dynamics faithfully. However, in this regime, the success probability still shows the expected trend of QA: a larger success probability upon increasing the effective annealing time by either increasing $p$ for a constant value of $\tau$ or vice-versa. This suggests that for these values of $\tau$ the circuit is implementing an effective QA Hamiltonian.
Thus, for a given $p$, the smallest available $\tau$ might not always be the optimal choice when practical considerations are taken into account. The regime of intermediate $\tau$ values can therefore be particularly interesting, as it enables longer effective annealing times to be achieved with shallower circuits, making this approach especially suitable for present-day NISQ devices. This regime of operation is similar to the ones explored in Refs.~\cite{zhou_quantum_2020,streif2020training,sack2021quantum,willsch_gpu-accelerated_2022,montanez2025toward}.
\begin{figure}[t]
    \centering
    \includegraphics[width=\linewidth]{Images/AQA_Const_Lines.png}
    \caption{Success probability of AQA with first-order approximation: variation with time step ($\tau$) and number of steps ($p$). Red lines denote constant total annealing time $\tau p$.}
    \label{fig:aqa-1}
\end{figure}
Lastly, beyond this region, i.e., for $\tau > 0.9$, the approximation degrades further, and the trajectory of the quantum circuit deviates significantly from that of quantum annealing. The corresponding circuit acts as a sequence of random quantum jumps in the Hilbert space, leading to a substantial decrease in the success probability.

\begin{figure}[!ht]
    \centering
    \begin{subfigure}[b]{\columnwidth}
        \centering
        \begin{tikzpicture}
            \node[inner sep=0pt] (img){\includegraphics[width=0.95\columnwidth]{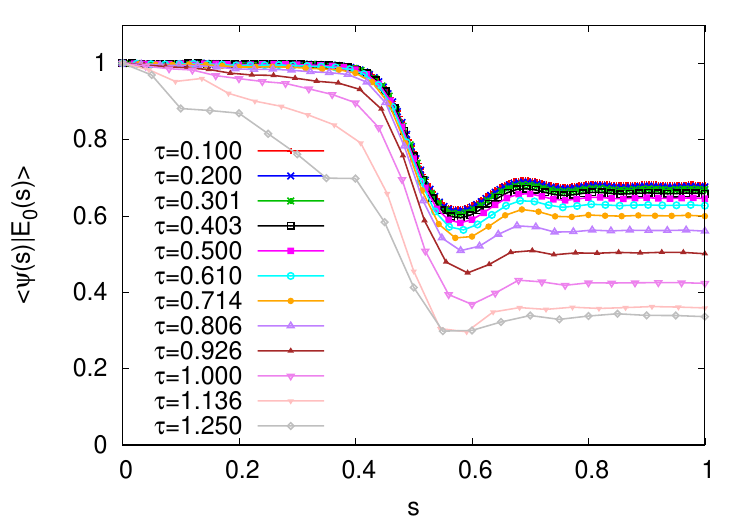}};
            \node[anchor=north west, font=\normalfont, xshift=-7pt, yshift=-4pt]
                at (img.north west) {(a)};
        \end{tikzpicture}
    \end{subfigure}

    \vspace{0.35cm}

    \begin{subfigure}[b]{\columnwidth}
        \centering
        \begin{tikzpicture}
            \node[inner sep=0pt] (img)
                {\includegraphics[width=0.95\columnwidth]{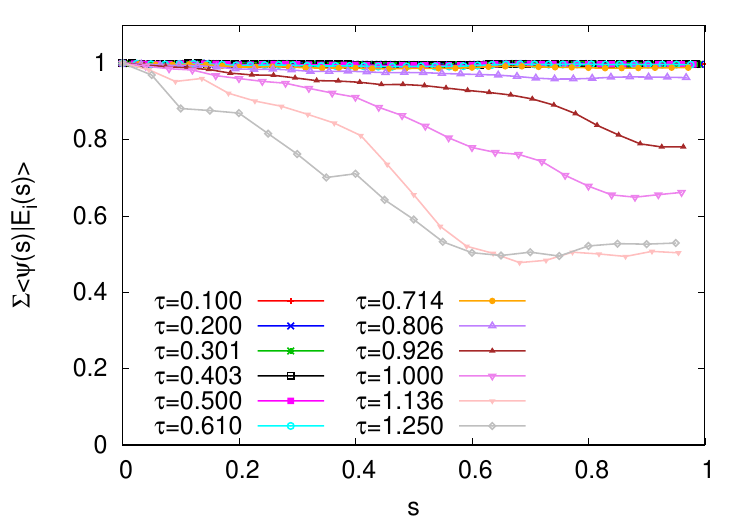}};
            \node[anchor=north west, font=\normalfont, xshift=-7pt, yshift=-4pt]
                at (img.north west) {(b)};
        \end{tikzpicture}
    \end{subfigure}

    \caption{Overlap of the instantaneous state of an 8-variable system evolving under AQA with the (a) ground state $E_0(s)$ and (b) the lowest three energy states of the instantaneous QA Hamiltonian $H(s)$ given by Eq.~(\ref{eq:qa-h}) for $t_a=25$ and various $\tau$.}
    \label{fig:aqa_instoverlap}
\end{figure}

Furthermore, Fig.~\ref{fig:aqa_instoverlap}(a) shows the overlap between the instantaneous state of this 8-variable problem evolving under AQA and the ground state $E_0(s)$ of the instantaneous QA Hamiltonian defined in Eq.~(\ref{eq:qa-h}), for $t_a=25$ and various values of $\tau$. For $\tau < 0.4$, the success probability ($\braket{\psi(1)|E_0(1)}$ in Fig.~\ref{fig:aqa_instoverlap}(a)) remains nearly constant. In this regime, the corresponding circuits effectively implement the same QA dynamics for the chosen annealing time. As $\tau$ increases, however, the success probability decreases. The larger-$\tau$ region can be further divided into two distinct regimes. First, for $0.4 \leq \tau \leq 0.9$, the system state closely tracks the ground state of the instantaneous QA Hamiltonian up to the anticrossing point (at $s=0.47$), after which it begins to deviate. This behavior indicates that, within this range, the AQA circuit still approximates an effective QA Hamiltonian. Second, for $\tau > 0.9$, the system state departs from the ground state already at small values of $s$, suggesting a complete breakdown of the connection to the intended QA dynamics.

Figure~\ref{fig:aqa_instoverlap}(b) further shows the sum of the overlaps of the instantaneous state with the ground, first excited, and second excited states of the instantaneous QA Hamiltonian, again for $t_a=25$ and various $\tau$. For $\tau < 0.9$, this sum remains close to one, indicating that the system’s evolution is largely confined to the lowest three energy levels. In contrast, for $\tau > 0.9$, the state acquires substantial overlap with higher excited states.

\section{Quantum Approximate Optimization Algorithm}
\label{sec:QAOA}
\subsection{Basic idea}
QAOA is a variational algorithm that employs a layered circuit ansatz structurally related to Eq.~(\ref{eq:discretized_qa}). Each layer introduces a pair of variational parameters, $\beta_i$ and $\gamma_i$, $i=1, \ldots, p$, which control the unitary operations generated by $H_I$ and $H_P$, respectively. By optimizing these $2p$ parameters according to the cost function given by the average energy of $H_P$, QAOA seeks a state with large overlap with the ground state of $H_P$ without relying on a continuous annealing trajectory. 

The quantum state produced by this circuit is optimized using classical routines to minimize the expected value of $H_P$,
\[
E(\vec{\beta}, \vec{\gamma}) = \bra{\psi(\vec{\beta}, \vec{\gamma})} H_P \ket{\psi(\vec{\beta}, \vec{\gamma})},
\]
where, 
\begin{equation}
    \ket{\psi(\vec{\beta}, \vec{\gamma})} = \prod_j e^{i\beta_j H_I} e^{-i\gamma_j H_P} \ket{+}^{\otimes N},
\end{equation}
with the uniform superposition state $\ket{+}^{\otimes N}$ serving as the initial state. We denote the optimized set of parameters obtained from the classical routine as
$(\vec{\beta^f},\vec{\gamma^f})$. The number of layers $p$ determines the expressive power of the ansatz: typically larger $p$ values allow obtaining parameters corresponding to states with larger overlap with the ground state of the problem Hamiltonian, but at the expense of increased circuit depth and classical optimization overhead.

In certain circumstances, ($\vec{\beta^f}$, $\vec{\gamma^f}$) are found to be close to their quantum annealing counterparts for specific annealing schedules, effectively resembling a QA trajectory. In general, however, QAOA treats these parameters as independent variational quantities, allowing the algorithm to explore trajectories that deviate from QA while potentially achieving a better overlap with the optimal solution for a given circuit depth. The performance of QAOA depends critically on the ability of the classical optimizer to identify parameter values ($\vec{\beta^f}$, $\vec{\gamma^f}$) that minimize the expectation value of $H_P$. In practice, this success is strongly influenced by the choice of the initial parameters $(\vec{\beta}^{i}, \vec{\gamma}^{i})$ used to initialize the optimization over the $2p$-dimensional parameter space. This naturally raises the question of which initial parameter values $(\vec{\beta}^{i}, \vec{\gamma}^{i})$ and how many layers $p$ lead to reasonable QAOA performance.

\subsection{Results}

\begin{figure}
    \centering
    \includegraphics[width=\linewidth]{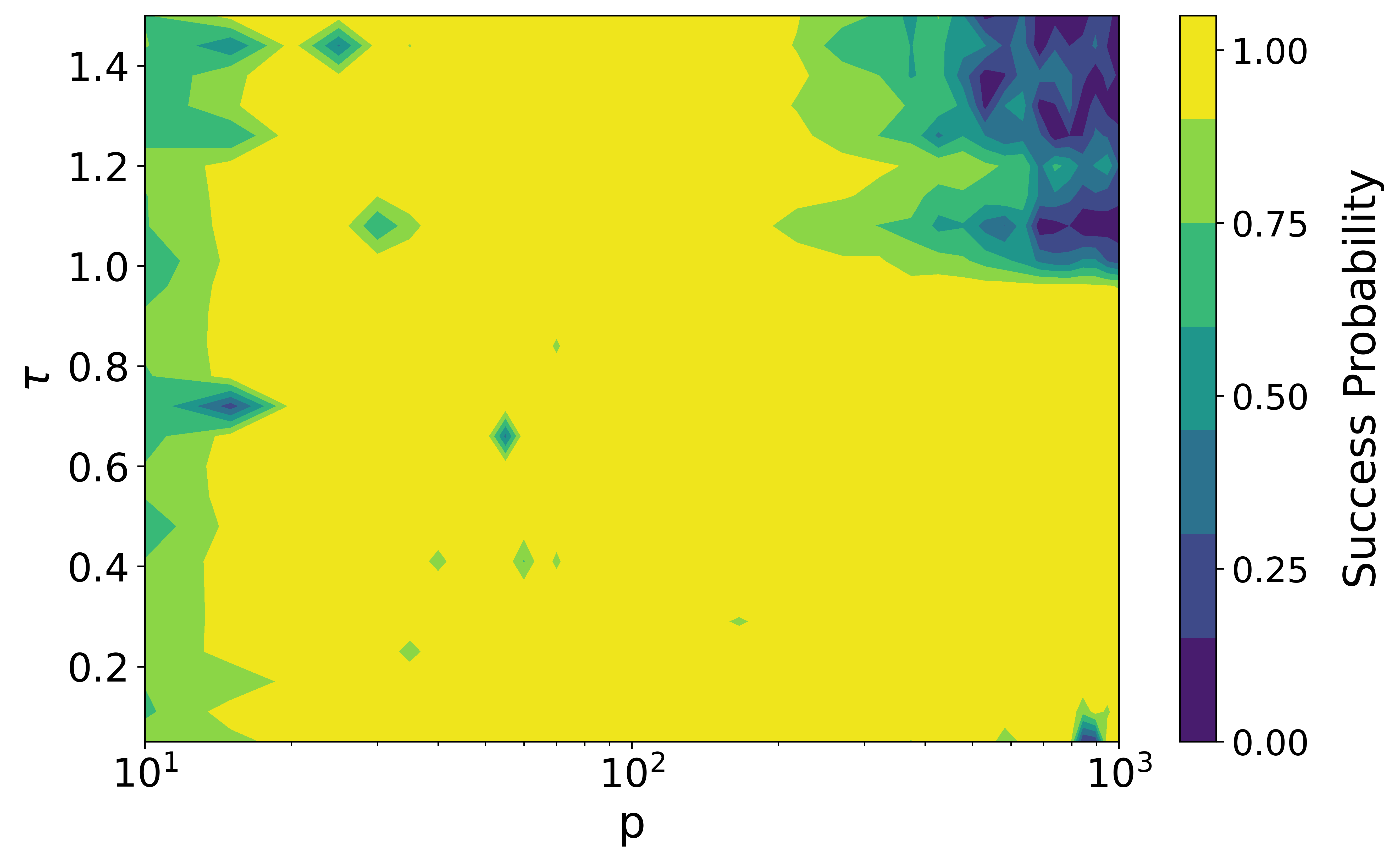}
    \caption{Success probability of QAOA initialized with AQA parameters: $\tau \in [0.05, 1.5]$, $p \in [10, 1000]$, capped at 8000 function evaluations.}
    \label{fig:aqa-qaoa-1}
\end{figure}

To investigate the above consideration, we again use the 8-variable 2-SAT instance examined for AQA in the previous section. The parameters used for obtaining the results in Fig.~\ref{fig:aqa-1} are chosen as the initial parameters $(\vec{\beta^i},\vec{\gamma^i})=\{(\tau A(j/p),\tau B(j/p)) ,\; j = 1,...,p\}$ for QAOA. The corresponding results for the success probability obtained using the BFGS optimizer are shown in Fig.~\ref{fig:aqa-qaoa-1} as a function of the initial parameters $\tau$ and $p$.



From Fig.~\ref{fig:aqa-qaoa-1}, it can be observed that for $\tau < 0.9$, the success probability is close to 1  except for small $p$. This behavior can be understood by noting that in this regime, the quantum circuit with initial parameters $(\vec{\beta^i},\vec{\gamma^i})$ resembles (effective) quantum annealing dynamics. This means that the corresponding state is confined to a low-energy subspace, which can dramatically simplify the energy landscape for the optimizer, as opposed to starting in a state involving higher energy levels.

As established in the previous section, Eq.~(\ref{eq:discretized_qa}) loses connection to QA for $\tau > 0.9$, and therefore the corresponding initialization of $(\vec{\beta^i}, \vec{\gamma^i})$ results in a state which involves higher energy levels. Such a state is less likely to provide a good starting point for the classical optimizer. Consequently, the optimization landscape can become highly nontrivial, and the classical optimizer may struggle to locate the global optimum. In principle, the starting state obtained using initial parameters in this regime might not be significantly different compared to that from a random initialization of the parameters, thereby causing the optimizer to be easily stuck in a local minimum. Nevertheless, as shown in Fig.~\ref{fig:aqa-qaoa-1}, for smaller values of $p$, the optimizer occasionally converges to parameter configurations that yield improved success probabilities. However, this behaviour is likely a consequence of the small problem size considered here and should not be expected to generalize to larger or more complex problem instances.

Thus, the strength of QAOA lies in its tendency to perform particularly well when initialized with physically meaningful parameters that produce a state having substantial overlap with the low-energy subspace of $H_P$. In such cases, the variational updates of the parameters $\vec{\beta}$ and $\vec{\gamma}$ might confine the evolved state to a subspace dominated by low-lying energy states, which significantly simplifies the optimization landscape. This can enhance the practical performance of the algorithm. 

\section{Evolving Hamiltonian Quantum Optimization}
\label{sec:EHQO}
\subsection{Basic idea}
So far, we have discussed approaches that borrow the ansatz and initial parameters from quantum annealing. In this section, we discuss the evolving Hamiltonian quantum optimization (EHQO) algorithm as a variational framework that explicitly incorporates QA-inspired guidance throughout the optimization process. More specifically, EHQO performs a multi-step optimization in which the Hamiltonian is adaptively updated at each stage, mimicking the progressive evolution of the instantaneous Hamiltonian in QA. This approach retains the flexibility of variational optimization while embedding QA-derived structure to guide the search toward high-quality solutions.

We start in the ground state of the initial Hamiltonian $H_{I}$ (Eq.~(\ref{eq:H_I})), i.e., the uniform superposition state, and apply a parameterized circuit, for instance, a QAOA ansatz, which, using the first intermediate Hamiltonian $H(s_1)$ as the target Hamiltonian, attempts to variationally find the parameters $\vec{\theta_1}$ that drive the initial state toward the ground state of $H(s_1)$. This is achieved by minimizing the cost function $C_1$, where
\begin{equation}
    C_j(\vec{\theta_j}) = \bra{\psi(\vec{\theta_j})} H(s_j) \ket{\psi(\vec{\theta_j})}, \;\; j=1,...,N_s.
\end{equation}

In the next step, the new parameters $\vec{\theta_2}$ are initialized to the values of the final parameters $\vec{\theta}_1^f$ from the previous step and are variationally optimized so that the state obtained from the second step has a good overlap with the ground state of $H(s_2)$. The method then continues iteratively: at each step, the state obtained in the previous stage serves as the starting point for the next variational stage, i.e., the parameters optimized in that earlier step initialize the next optimization, ideally enabling the algorithm to obtain a state having a good overlap with the ground state of the Hamiltonian for the next step. This process continues until the problem Hamiltonian is reached.

It is worth noting that this method is not specific to any particular ansatz for the parameterized circuit, i.e., EHQO can be applied to any parameterized quantum circuit. This flexibility is especially valuable in the NISQ era, where the choice of ansatz must conform to hardware constraints. A drawback of EHQO, however, is the need to perform a variational optimization at every evolution point, which incurs an $N_s$-fold increase in computational cost. This overhead can be interpreted as the first $N_s-1$ optimization steps serving primarily to provide meaningful initial parameters for the final step. In this work, we employ the QAOA ansatz. In practice, this requires choosing the number of QAOA layers $p$, which may optionally be step dependent ($p_j$).

\subsection{Practical considerations}
The performance of EHQO can be expected to depend on the choice of certain tunable parameters. These include the following.
\begin{itemize}
    
    \item \textbf{Choice of ansatz}: For any variational algorithm, the choice of ansatz plays a crucial role in determining its ability to find the optimal state. In the present study, we employ the QAOA ansatz, for which the expressive power is characterized by the number of variational parameters. While increasing this number generally improves the representational capacity of the ansatz, allowing it to more accurately approximate the ground state of the Hamiltonian $H(s_j)$ at each step, it also leads to a more complex optimization landscape. As a result, the optimizer is more likely to become trapped in local minima, potentially hindering the search for the global minimum.

    Since EHQO is inherently step-based, updating both the target Hamiltonian and the variational parameters at each stage, we can additionally choose to adapt the circuit depth at each step.

    \item \textbf{Number of steps ($N_s$):} EHQO involves $N_s$ Hamiltonians based on the quantum annealing Hamiltonian. Increasing $N_s$ results in smoother Hamiltonian transitions, closely resembling a QA-like evolution. However, excessive steps make the process computationally intensive, like quantum annealing itself. Conversely, too few steps in the process risk an abrupt change in the ground states of the consecutive Hamiltonians, which can undermine the appeal of this approach. It can therefore be expected that the critical region for EHQO coincides with that of quantum annealing, namely, the region in the vicinity of the anticrossing.
    
    \item \textbf{Classical optimizer:} The choice of a classical optimizer plays an important role in determining both the efficiency and reliability of the variational procedure. Different optimizers exhibit distinct trade-offs in terms of convergence speed, robustness to noise, and sensitivity to the structure of the optimization landscape. Gradient-based optimizers can converge rapidly when reliable gradient information is available, but may struggle in regions with flat landscapes or vanishing gradients, or become trapped in local minima. In the context of EHQO, where the optimization landscape evolves across stages, and the number of variational parameters can grow, the choice of optimizer can significantly impact overall performance and scalability.

\end{itemize}

Since QA starts in the uniform superposition state, which is the ground state of $H_I$, we use this initial state for EHQO, and initialize the parameter set to $(\vec{\beta}_1^i,\vec{\gamma}_1^i) = (\vec{0},\vec{0})$. However, it was found that while this choice performed well for the Nelder-Mead optimizer, it consistently caused the BFGS optimization to stall. This behavior can be attributed to the fact that small parameter perturbations around this highly symmetric initialization can lead to nearly vanishing estimates of the cost-function gradients, leaving BFGS without a meaningful search direction. To address this, we introduce a small perturbation to the initial parameters for the BFGS study, setting them to $\epsilon$. We experimented with $\epsilon$ values of $10^{-2}$, $10^{-4}$, and $10^{-6}$ and evaluated the resulting success probabilities. Based on these observations, we select $\epsilon = 10^{-2}$ for our simulations. This modification ensures that the initial gradients are non-zero, allowing BFGS to explore the energy landscape without being stuck. 
\subsection{Results}

\begin{figure}[!ht]
    \centering
    \begin{subfigure}[b]{\columnwidth}
        \centering
        \begin{tikzpicture}
            \node[inner sep=0pt] (img)
                {\includegraphics[width=0.95\columnwidth]{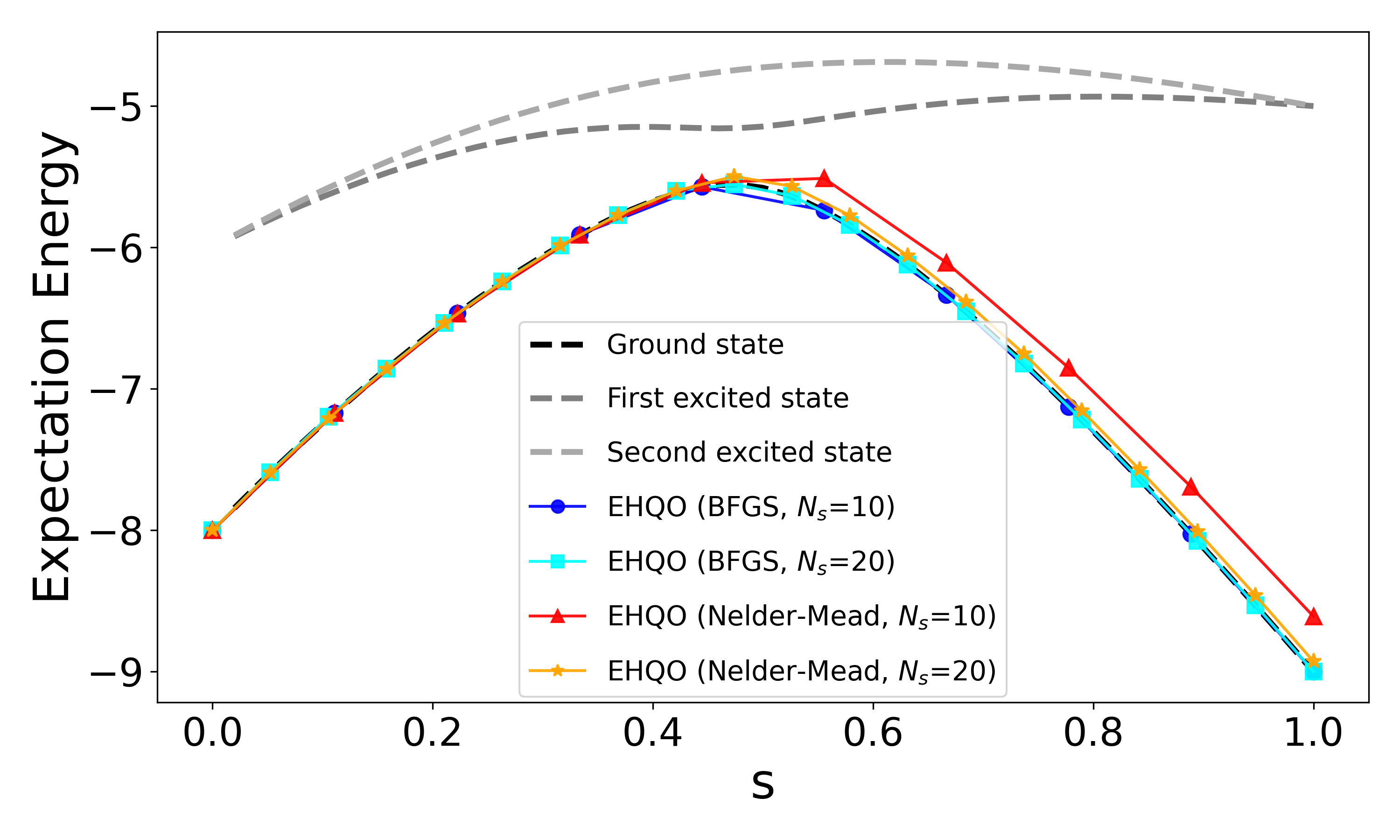}};
            \node[anchor=north west, font=\normalfont, xshift=-7pt, yshift=-4pt]
                at (img.north west) {(a)};
        \end{tikzpicture}
    \end{subfigure}

    \vspace{0.35cm}

    \begin{subfigure}[b]{\columnwidth}
        \centering
        \begin{tikzpicture}
            \node[inner sep=0pt] (img)
                {\includegraphics[width=0.95\columnwidth]{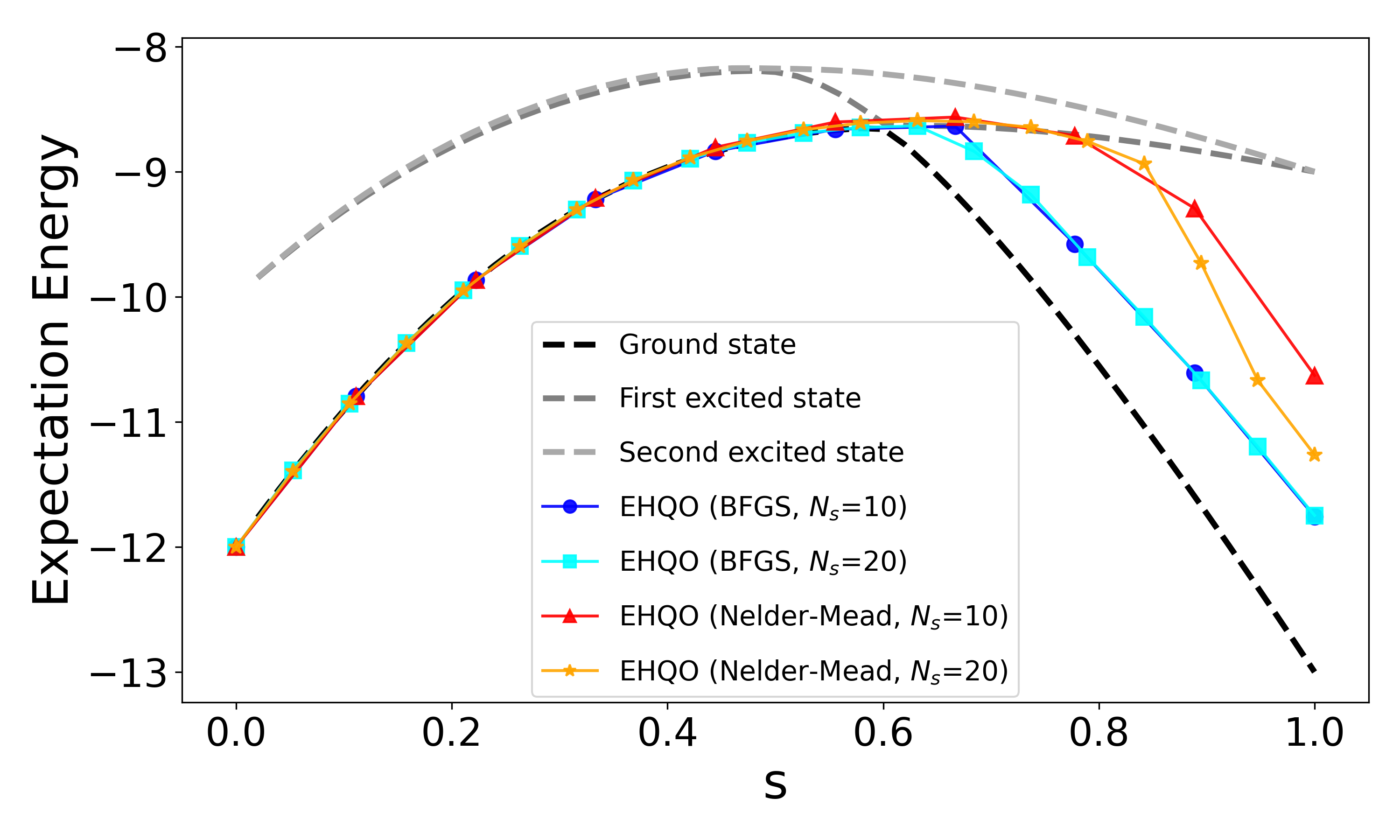}};
            \node[anchor=north west, font=\normalfont, xshift=-7pt, yshift=-4pt]
                at (img.north west) {(b)};
        \end{tikzpicture}
    \end{subfigure}

    \caption{Expectation value of $H(s_j)$ with respect to the final state at each step $j$ under the EHQO algorithm using the QAOA ansatz with $p=25$ and $N_s=10,20$.}
    \label{fig:ehqo_instance}
\end{figure}

To gain insight into the workings of the EHQO algorithm, we analyze an instance of 8- and 12-variable problems each. Figure~\ref{fig:ehqo_instance} shows the expectation value of the Hamiltonian $H(s_j)$ with respect to the state obtained during the EHQO evolution using a QAOA ansatz with fixed depth $p = 25$ and $N_s = 10$ and $20$ Hamiltonians, together with the energy spectrum of the quantum annealing Hamiltonian $H(s)$. As shown in Fig.~\ref{fig:ehqo_instance}(a), for the 8-variable system, the expectation energy closely follows the ground-state energy throughout the evolution, demonstrating that using the state obtained from the previous step as the initial state for the next step yields good results. In contrast, Fig.~\ref{fig:ehqo_instance}(b) shows that, for the larger 12-variable system, the expectation energy closely follows the ground-state energy of the intermediate Hamiltonians only during the early stages of the evolution. Near the anticrossing region, however, it begins to deviate toward excited-state energies. This behavior suggests that this region may act as a bottleneck for the overall performance of the algorithm.

\begin{figure}[!ht]
    \centering
    \begin{subfigure}[b]{\columnwidth}
        \centering
        \begin{tikzpicture}
            \node[inner sep=0pt] (img){\includegraphics[width=0.95\columnwidth]{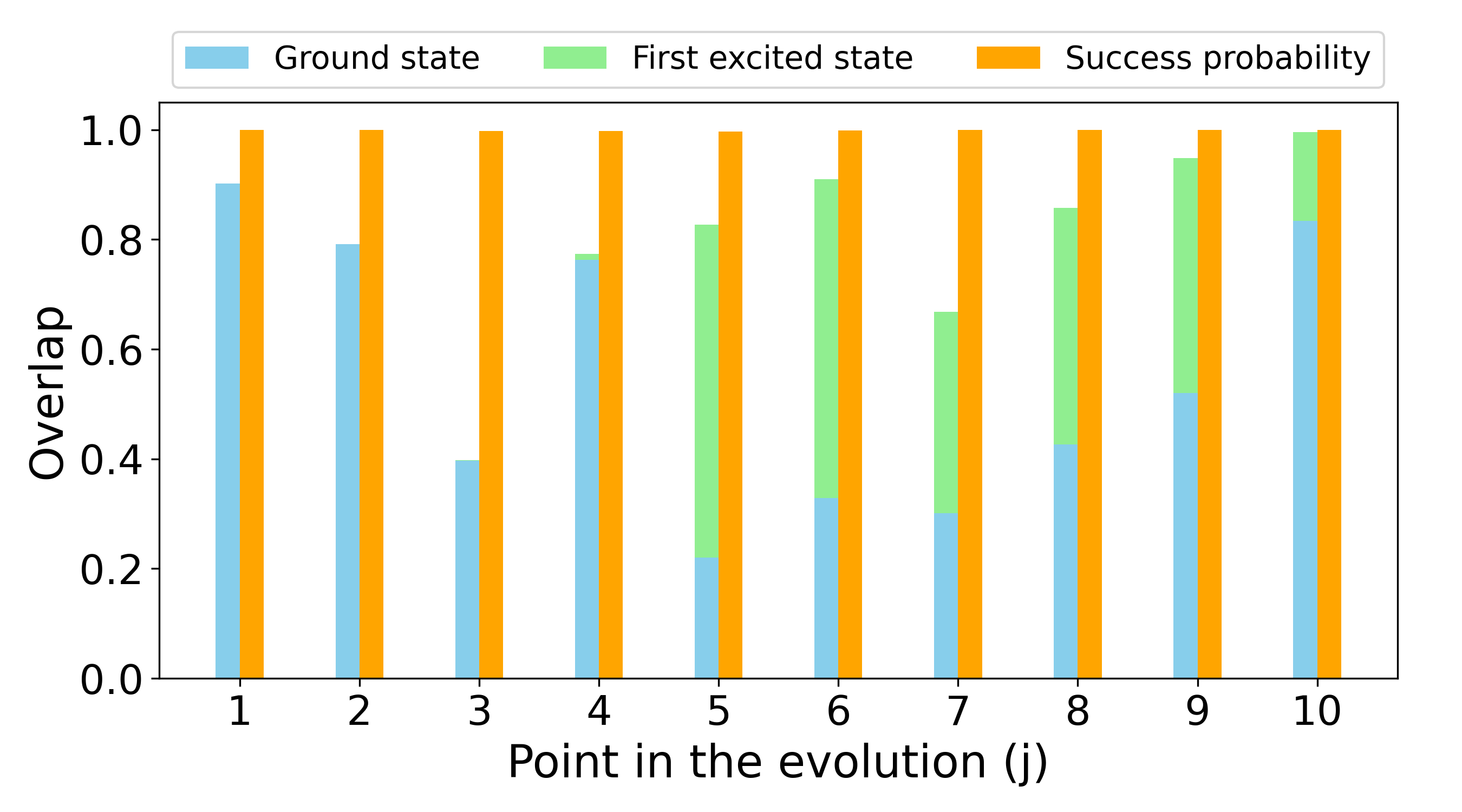}};
            \node[anchor=north west, font=\normalfont, xshift=-7pt, yshift=-4pt]
                at (img.north west) {(a)};
        \end{tikzpicture}
    \end{subfigure}

    \vspace{0.35cm}

    \begin{subfigure}[b]{\columnwidth}
        \centering
        \begin{tikzpicture}
            \node[inner sep=0pt] (img)
                {\includegraphics[width=0.95\columnwidth]{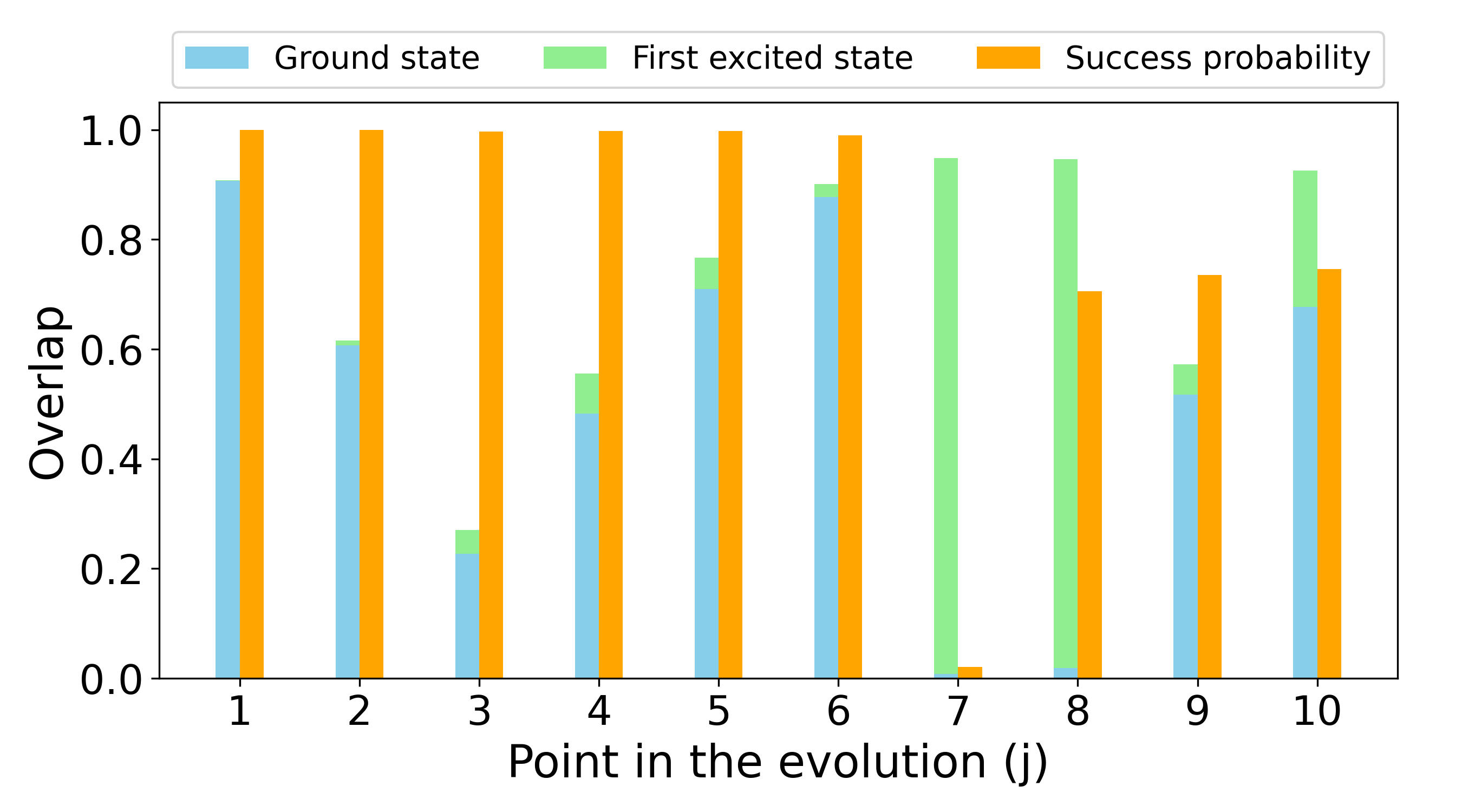}};
            \node[anchor=north west, font=\normalfont, xshift=-7pt, yshift=-4pt]
                at (img.north west) {(b)};
        \end{tikzpicture}
    \end{subfigure}

    \caption{Overlaps in EHQO for an 8 and 12 -qubit Hamiltonian of the initial wavefunction (i.e., using parameters from the previous optimization process) with the ground state and the first excited state of the intermediate Hamiltonians. Also depicted are the final success probabilities after optimization at each point.}
   \label{fig:8-ehqo-overlaps}
\end{figure}

Further insight is provided by Fig.~\ref{fig:8-ehqo-overlaps}, which shows the overlap of the initial state at each step, namely the final state from the previous step, with the ground and first excited states of the corresponding intermediate Hamiltonian. The figure also shows the overlap of the state obtained after optimization at each step with the ground state, i.e., the success probability. As seen in Fig.~\ref{fig:8-ehqo-overlaps}(a) for the 8-variable problem, the initial state inherited from the previous step typically has a substantial overlap with the low-lying energy states of the current Hamiltonian, making it a suitable starting point for the next optimization stage.
From Fig.~\ref{fig:8-ehqo-overlaps}(b) for the 12-variable system, the final state obtained at $s_6$, which has a large overlap with the ground state of $H(s_6)$, has essentially zero overlap with the ground state of $H(s_7)$ and instead a significant overlap with its first excited state. This is a manifestation of crossing the point of a Landau-Zener-like anticrossing in the corresponding QA Hamiltonian, where the ground states before and after the point of anticrossing are orthogonal. Consequently, after optimization at $s_7$, the resulting state achieves only a small overlap with the ground state of $H(s_7)$. This suggests that passing through this region might be a bottleneck for the algorithm. Nevertheless, for this particular instance, the optimizer is still able to find states with substantial overlap with the ground states of the following Hamiltonians in the subsequent steps.

In summary, the EHQO algorithm can yield good results by using the final state obtained at the previous step as the initial state for the next step. At the last step, EHQO can be viewed as a systematic search for a good initial state (parameters) through a sequence of optimization processes. However, a region where the ground states of the consecutive intermediate Hamiltonians differ significantly can form a bottleneck for the overall algorithmic performance. One possible mitigation strategy is to increase the number of intermediate Hamiltonians in this critical region, although this approach comes at the cost of increased computational overhead.

\section{Comparative performance analysis}
\label{sec:PerformanceComparison}
Having discussed the basic ideas underlying the AQA, QAOA, and EHQO algorithms and examined their dynamical behavior on representative problem instances, we now turn to a systematic comparison of their performance. To this end, we benchmark these algorithms on ensembles of hard 2-SAT problem instances. We first analyze their performance under different algorithmic settings for the 8- and 12-variable problem sets and subsequently investigate how their performance scales with system size. While the analyses of individual instances presented earlier provide useful intuition about the underlying mechanisms, the ensemble-level results reported here offer a statistically more robust assessment of algorithmic performance. All simulations are carried out using the Qiskit emulator~\cite{qiskit2024} or JUQCS~\cite{de_raedt_massively_2007,de_raedt_massively_2019}, and results are reported for both a gradient-based optimizer (BFGS) and a gradient-free optimizer (Nelder–Mead).

\subsection{8- and 12-variable problem sets}

\begin{figure}[!ht]
    \centering
    \begin{subfigure}[b]{\columnwidth}
        \centering
        \begin{tikzpicture}
            \node[inner sep=0pt] (img)
                {\includegraphics[width=0.95\columnwidth]{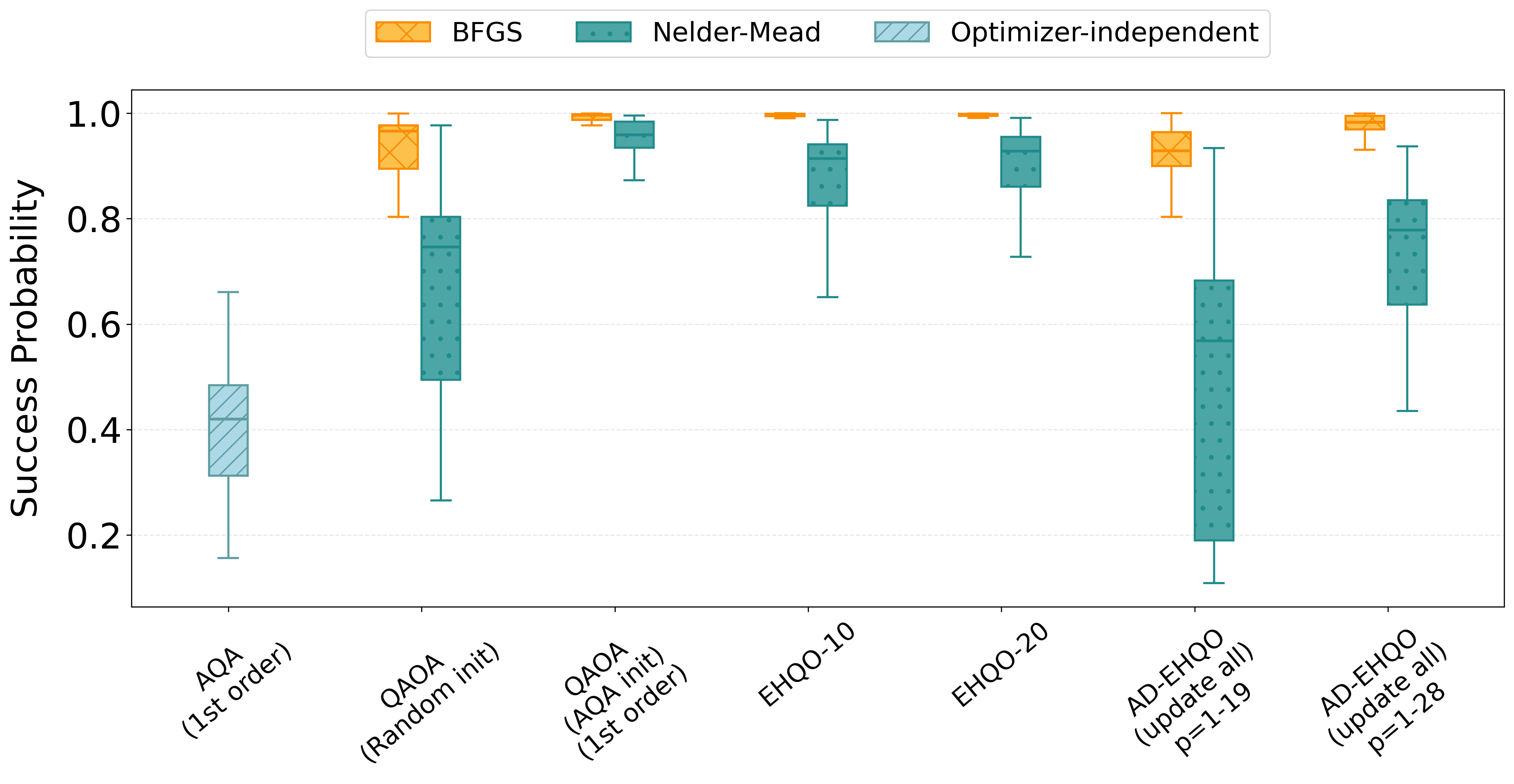}};
            \node[anchor=north west, font=\normalfont, xshift=-7pt, yshift=-4pt]
                at (img.north west) {(a)};
        \end{tikzpicture}
    \end{subfigure}

    \vspace{0.35cm}

    \begin{subfigure}[b]{\columnwidth}
        \centering
        \begin{tikzpicture}
            \node[inner sep=0pt] (img)
                {\includegraphics[width=0.95\columnwidth]{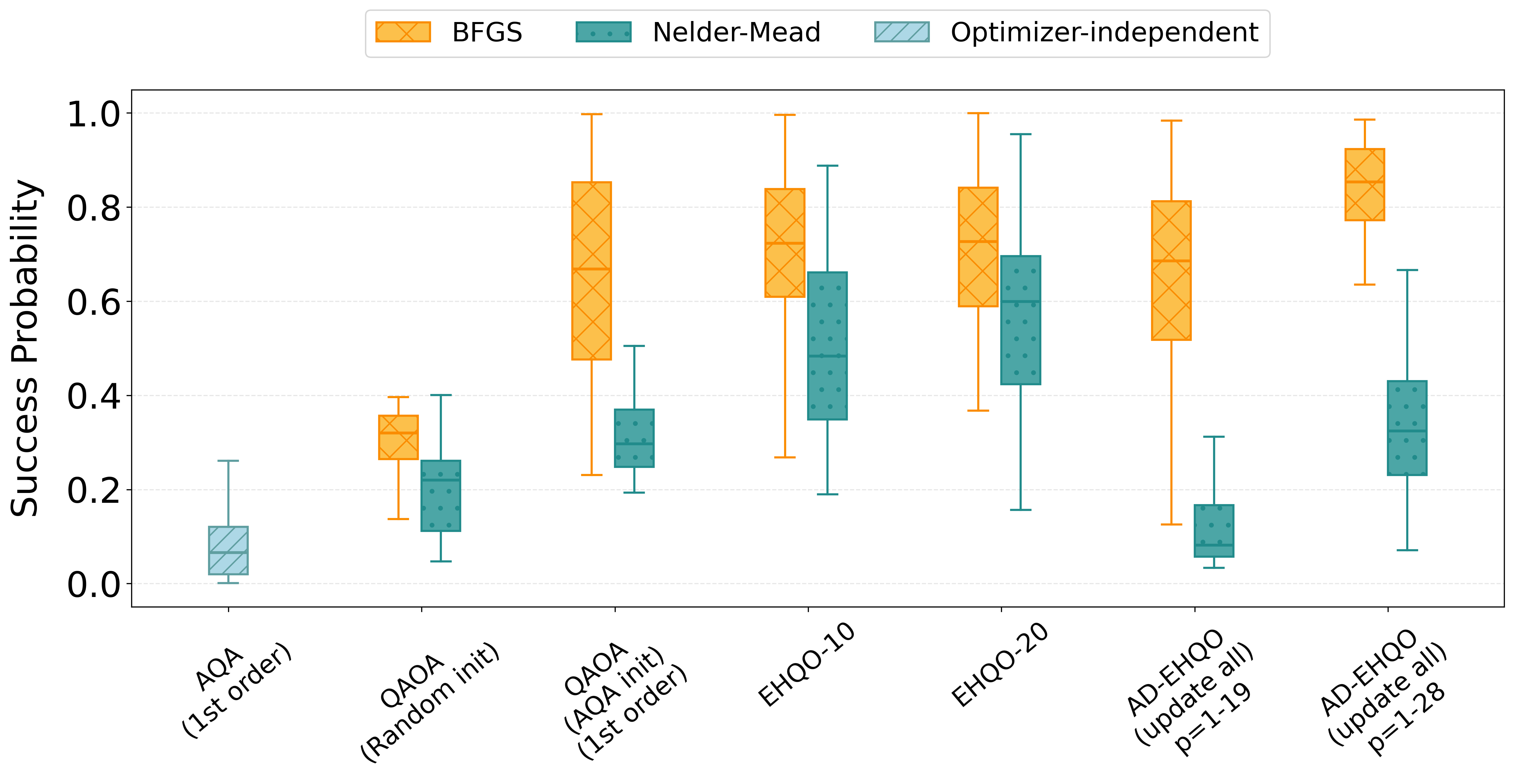}};
            \node[anchor=north west, font=\normalfont, xshift=-7pt, yshift=-4pt]
                at (img.north west) {(b)};
        \end{tikzpicture}
    \end{subfigure}

    \caption{Quartiles of success probabilities for different types of EHQO for (a) 8-qubit and (b) 12-qubit Hamiltonian using both BFGS and Nelder-Mead.}
    \label{fig:quartiles}
\end{figure}

We begin by comparing the performance of the considered algorithms on the 8- and 12-variable problem sets introduced above. The results are reported in terms of success probability across different algorithmic settings. For EHQO, we consider two fixed-depth variants with $p=25$ at each step and $N_s=10$ (EHQO-10) and $N_s=20$ (EHQO-20), as well as an adaptive-depth version (AD-EHQO) with $N_s=10$. In the adaptive variant, the depth of the QAOA ansatz is gradually increased from one step to the next, with the number of layers growing from 1–19 (in increments of 2) and from 1–28 (in increments of 3), while reoptimizing all parameters at each stage. These results are compared with those obtained using AQA with parameters $p=25$, $\tau=0.5$, and a linear annealing schedule, as well as with QAOA ($p=25$) using either random initializations or QA-inspired parameter initializations identical to those used in AQA. The corresponding performance statistics are shown in Fig.~\ref{fig:quartiles}, where we report the quartiles of the success probabilities for the 8- and 12-variable problem sets.

The first observation from the figure is that AQA with parameters fixed to $\tau=0.5$ and $p=25$ performs rather poorly as the effective annealing time $\tau p$ is not sufficiently large in this otherwise QA-like regime. Secondly, the Nelder--Mead optimizer performs consistently worse than the BFGS optimizer for these problems. The reasons underlying this discrepancy lie beyond the scope of the present work and require a dedicated investigation. Consequently, the discussion that follows is based on results obtained using the BFGS optimizer.

From Fig.~\ref{fig:quartiles}(a), we observe that for the 8-variable 2-SAT instances, the median success probability is close to 1 for all the hybrid methods. This uniformly high performance is due to the relatively small size of these instances. To better distinguish the behavior of the different algorithms, we therefore turn our attention to the 12-variable results presented in Fig.~\ref{fig:quartiles}(b). We find that all variants of EHQO achieve higher median success probabilities than QAOA with random initializations. However, the QAOA ansatz based on the first-order product-formula approximation of QA exhibits a performance comparable to EHQO. Note that at its last step, EHQO differs from QAOA only in the values of the initial parameters. In the former, the initial parameters for the last step are obtained systematically through the succession of previous optimization processes, while in the latter, the initial parameters are chosen freely. As evident from these results, initial parameters derived from QA serve as a better starting point compared to random initializations.


Within the EHQO block, the fixed-depth variants (EHQO-10 and EHQO-20) show competitive performance. AD--EHQO, where the circuit depth grows during the evolution, shows greater variability. Small final depths perform worse than larger final depths, showing a comparable performance to fixed depth EHQO. This highlights an advantage of AD--EHQO: it naturally adapts to the observation that shallow circuits are sufficient in the early stages of the evolution, while at later stages require a deeper ansatz. Consequently, AD--EHQO can offer greater versatility by balancing computational resources and performance. 

Across all methods, the success probabilities decrease when moving from 8 to 12 variables. This is expected due to the rapidly growing Hilbert space and the increasing complexity of the spectral features. This trend naturally motivates us to look at how the proposed algorithm scales with increasing problem size.

\subsection{Scaling behavior of success probability}

Studying the scaling behavior is an essential step in evaluating the efficiency of an algorithm as problem sizes grow. Based on previous results, we focus on the algorithmic performance using the BFGS optimizer.

\begin{figure}
    \centering
    \includegraphics[width=0.8\linewidth]{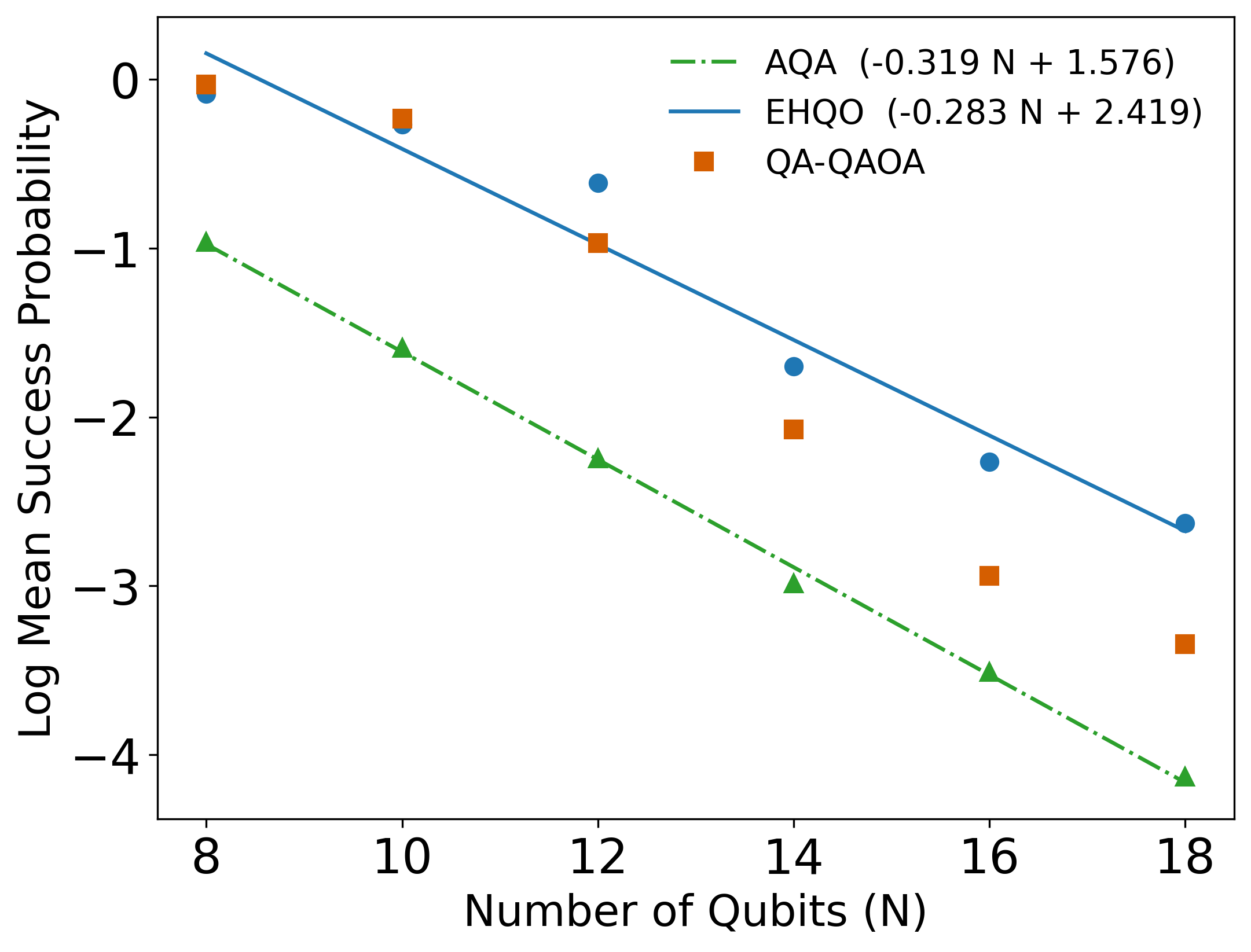}
    \caption{Scaling of mean success probability for 100 instances of 8-18 qubit 2-SAT problems for AQA (green triangles) and EHQO (blue circles).}
    \label{fig:aqa-qaoa-scaling}
\end{figure}

To this end, we compare the scaling performance of AQA and QAOA with that of fixed-depth EHQO using $N_s = 10$. For AQA, we once again use parameters $\tau=0.5$ and $p=25$, and a linear schedule (see Eq.~(\ref{eq:discretized_qa})). We initialize $(\vec{\beta}, \vec{\gamma})$ for QAOA using the same parameters as AQA. Furthermore, for the initial parameters in the first step of EHQO, we also use the first-order product-formula approximation of QA with $p = 25$. Note that this initialization strategy for fixed-depth EHQO differs from the earlier choice of setting all parameters to $10^{-2}$.



We used the in-house simulator JUQCS to obtain the scaling results. Furthermore, to partially tackle the issue of increasing computational costs with increasing problem sizes, we impose limits on the number of JUQCS function calls, setting a cap of 50,000 for the final optimization points while restricting intermediate runs to 10,000 calls. Using this strategy, we analyze the scaling of the mean success probability across 100 instances of 2-SAT problems ranging from 8 to 18 variables, shown in Fig. \ref{fig:aqa-qaoa-scaling}. We find that the success probability obtained from AQA decreases exponentially with increasing problem size, with an exponent of -0.319. The values of the success probabilities are greatly improved further by QAOA. Nevertheless, the scaling of success probabilities from QAOA still seems to be exponentially decaying. However, the scaling exponent cannot be reliably determined due to the limited number of data points. The scaling trend appears to be parallel to that for AQA. Moreover, the success probabilities are further improved by EHQO with an apparent smaller scaling exponent $-0.283$. This indicates that EHQO manages to arrive at a better initialization at the final step as compared to the AQA initialization for QAOA.

\section{Conclusion}
\label{sec:conclusion}

In this work, we studied quantum optimization algorithms inspired by the quantum annealing algorithm tailored for the NISQ era. We examined three such algorithms - approximate quantum annealing (AQA), the well-known and studied QAOA, and the so-called evolving Hamiltonian quantum optimization (EHQO), through numerical simulations for hard 2-SAT problems with 8 to 18 variables.  

Our results demonstrate that, within the discretized and parameter-tunable extension of quantum annealing, i.e., AQA, there exist parameter regimes that reproduce annealing-like behavior while requiring fewer resources than those needed for faithfully implemented quantum annealing. Furthermore, the parameters obtained from AQA were found to provide an effective initialization strategy for QAOA, with such informed initializations yielding significantly improved performance compared to AQA, as well as to QAOA starting from random initializations for large system sizes.

We further studied EHQO, in which optimization is carried out sequentially over a set of intermediate Hamiltonians derived from a quantum annealing Hamiltonian. The motivation is that, by following a quantum annealing–like trajectory, the algorithm guides the system’s state toward states with low expectation value of the problem Hamiltonian, thereby reducing the likelihood of becoming trapped in local minima, as can occur in single-step variational algorithms. Alternatively, EHQO can be interpreted as a strategy for generating high-quality initial parameters at each step via the sequential optimization of parameters obtained from the preceding step.

For the problem instances considered, EHQO achieves higher average success probabilities than AQA and QAOA at comparable circuit depths. The scaling of the average success probability for EHQO exhibits a smaller exponent than that obtained for AQA, while QAOA shows scaling behavior similar to AQA when initialized with parameters for AQA. These observations suggest that gradually modifying the cost Hamiltonian can indeed facilitate the variational optimization process. Beyond its performance advantages, EHQO is also conceptually flexible, i.e., although our implementation employs QAOA ansatz as the variational circuit, the framework itself is ansatz-independent and can, in principle, be combined with any variational quantum circuit.

Taken together, these results suggest that incorporating annealing-inspired structure into quantum computing algorithms can be beneficial under realistic resource constraints. The methods studied here remain heuristic and are evaluated on small problem sizes. Nevertheless, they provide insight into how ideas from quantum annealing can be adapted and integrated into hybrid quantum–classical algorithms, offering practical strategies for near-term quantum optimization.

\section*{Acknowledgements}
The authors gratefully acknowledge the Gauss Centre for Supercomputing e.V. (www.gauss-centre.eu) for funding this project by providing computing time through the John von Neumann Institute for Computing (NIC) on the GCS Supercomputer JUWELS~\cite{JUWELS} at the Jülich Supercomputing Centre (JSC). R.S. and V.M. acknowledge support from the project JUNIQ funded by the German Federal Ministry of Education and Research (BMBF) and the Ministry of Culture and Science of the State of North Rhine-Westphalia (MKW-NRW). V.M. also acknowledges support from the project EPIQ funded by MKW-NRW.

\bibliography{references,references_1}
\end{document}